\documentclass[aps,prb,reprint,superscriptaddress]{revtex4-2}

\usepackage{braket}
\usepackage{amsmath, amsfonts, amssymb}
\usepackage{graphicx}
\usepackage{textcomp}
\usepackage{gensymb}
\usepackage{booktabs}

\usepackage{multirow}

\usepackage[caption=false]{subfig}
\newcommand{\phantomsubfloat}[1]{
    {
        \captionsetup[subfigure]{labelformat=empty}
        \subfloat[][]{#1}
    }%
}

\usepackage{times}
\usepackage{verbatim}

\usepackage[bbgreekl]{mathbbol}
\DeclareSymbolFontAlphabet{\mathbbm}{bbold}
\DeclareSymbolFontAlphabet{\mathbb}{AMSb}%

\usepackage[colorlinks=true,
            linkcolor=blue,
            anchorcolor=red,
            citecolor=red]{hyperref}
\usepackage{hypernat} 
\usepackage[capitalize]{cleveref}

\begin{document}


\title{Group theoretical and \emph{ab-initio} description of color center candidates in fluorographene}


\author{M. S. Tacca}
\email[]{marcos.tacca@uni-ulm.de}
\affiliation{Institut für Theoretische Physik and IQST, Albert-Einstein-Allee 11, Universität Ulm, D-89081 Ulm, Germany.}

\author{M. B. Plenio}
\email[]{martin.plenio@uni-ulm.de}
\affiliation{Institut für Theoretische Physik and IQST, Albert-Einstein-Allee 11, Universität Ulm, D-89081 Ulm, Germany.}


\date{\today}

\begin{abstract}

We present a group theoretical and \emph{ab-initio} analysis of lattice point defects in fluorographene, with a focus on neutral and negative $\text{V}_{\text{CF}}$ vacancies. 
By using a combination of density functional theory calculations  and group theory  analysis, we investigate the many-body configurations  of the defects and calculate the vertical absorption and zero-phonon line energies of the excited states and their dependence with strain. 
The description of the defects is extended by computing their formation energy, as well as further relevant parameters as the Jahn-Teller energy for neutral $\text{V}_{\text{CF}}$ and the zero field splitting for negative $\text{V}_{\text{CF}}$ vacancies.
Based on our results, we discuss possible quantum applications of these color centers when coupled to mechanical oscillation modes of the hosting two-dimensional material.
The symmetry and active orbitals of the defects exhibit a parallelism with those of the extensively studied NV centers in diamond. In this context, the studied defects emerge as interesting candidates for the development of two-dimensional quantum devices based on fluorographene.

\end{abstract}


\maketitle

\section{INTRODUCTION \label{sec:Introduction}}

Point defects are of increasing interest in the fields of quantum information and sensing due to their potential applications, among which are the promising NV center technologies \cite{Wrachtrup2006,Doherty2013,Wu2016}. 
By coupling the localized states introduced by color centers with mechanical oscillation modes, hybrid quantum devices with long-range interactions mediated by phonons can be fabricated through appropriate design \cite{Lee2017,Kepesidis2013,Ramos2013,Bennett2013c,Golter2016a}. 
The introduction of color centers in two-dimensional (2D) materials is particularly promising for the continuously accelerated development of quantum technologies. %

Two-dimensional resonators can be mechanically coupled with cavities through opto-thermal, electromagnetic, or further interactions \cite{Dolleman2018,Steeneken2021,Pokharel2022}.    
The dynamics of 2D membranes and other micro- and nano-devices have been widely studied for their potential applications in quantum and mass sensors, quantum simulators, and nanophotonics \cite{Steeneken2021,Burek2012,Chan2011,Hoese2020}.
Because color centers in 2D structures lie naturally on the surface of the material, high sensitivity to the environment is expected \cite{Abdi2018}. %
Various materials, including graphene \cite{Fandan2020,Barton2011,DeAlba2016a,Verbiest2021}, MoS${}_{\text{2}}$ \cite{Jiang2020,Castellanos-Gomez2013}, hexagonal boron nitrite (h-BN) \cite{Li2020,Abdi2017} and others \cite{Castellanos-Gomez2015,Steeneken2021,Blundo2021} have been studied as candidates for 2D systems.    
In particular, h-BN, a wide-band insulator that can host color centers \cite{Tran2016,Vaidya2023}, has been proposed as a platform for quantum simulation and ultra-sensitive force detection \cite{Li2020,Abdi2017,Abdi2018a,Gong2023}.   

In this work, we explore the potential of defect-bearing fluorographene \cite{Zboril2010,Nair2010} as a platform for the realization of hybrid quantum devices. 
Fluorographene (FG) is a stoichiometric 2D derivative of graphene, in which one fluorine atom is bonded to each carbon atom. 
This material has been used for a variety of applications, including electrochemical sensors, batteries, and electrocatalysis, as well as electronic applications such as transistors and solar cells  \cite{Chronopoulos2017}. 
A key characteristic of FG is that the carbon atoms exhibit $sp_3$ hybridization instead of the $sp_2$ one found in graphene. As a result, the electronic properties of FG are closer to those of diamond than to those of graphite. 
In fact, the structure of FG is similar to the fluorine-terminated (111) diamond surface, which has been  proposed as a suitable candidate for the implementation of a quantum simulator at room temperature \cite{Cai2013}. 
The application of polarized nuclear spins in quantum simulators is an active research field, in particular for the previously mentioned h-BN based systems \cite{Gao2022,Tabesh2023}.

Although it has been well established that FG presents a large band gap, its precise value has been a longstanding issue that appears to have been clarified only recently \cite{Hruby2022}.
Initial measurements suggested  a band gap larger than $3$ eV \cite{Nair2010}, and latter measurements yielded a value of $3.8$ eV, consistent with the first results \cite{Jeon2011}. 
Additional photoluminescence emission peaks have been observed at $3.56$ \cite{Mazanek2015}  and $3.65$ eV \cite{Mazanek2015,Jeon2011},  with the latter being attributed to phonon-assisted radiative recombination.
On the theoretical field, the initial density functional theory (DFT) \cite{Hohenberg1964} calculations  at the local density approximation (LDA) and generalized gradient approximation (GGA) theory levels 
resulted in predicted band gap values close to $3$ eV \cite{Zhou2010,Klintenberg2010,Leenaerts2010,Karlicky2012,Samarakoon2011}, 
in excellent agreement with the experimental measurements.
However, more refined calculations including the exact exchange interaction through the hybrid screened functional (HSE) predicted a larger  band gap of ${\approx}5$ eV \cite{Karlicky2012,Karlicky2013}.   

Additional calculations incorporating electron-electron interactions via Green's function methods (GW) on top of either LDA or GGA to further improve the description of the electronic structure, led to a predicted band gap of about  $7.5$ eV \cite{Samarakoon2011,Leenaerts2010,Karlicky2013,Klintenberg2010}. 
The inclusion of electron–hole interactions through the Bethe–Salpeter equation (BSE-GW), one of the most advanced methods beyond DFT, partially cancels the electron–electron interactions and results in predicted band gap values between $5.4$ \cite{Samarakoon2011} and $5.65$ eV \cite{Hruby2022}. 
It is worth noting that the latter  values are in agreement with the results obtained via the HSE method, which is computationally less demanding.

The discrepancies between the measured and calculated values of the band gap have been tentatively linked to midgap states  resulting from  defects in the material \cite{Leenaerts2010,Samarakoon2011}. A combined experimental and theoretical study has confirmed this hypothesis, showing that the  band gap value is in  agreement with previously reported BSE-GW results \cite{Hruby2022}.
The longstanding FG bandgap conundrum highlights  the importance of characterizing defects in materials. 
However, most theoretical works on FG have primarily focused on improving the accuracy of band gap predictions for the pristine material. Thus, the calculation of defects is often relegated to a secondary place \cite{Samarakoon2011,Karlicky2013,Wei2013}, or analyzed at the GGA level of the theory, which strongly underestimates the band gap \cite{Li2021}. 

In this work,  we investigate the electronic structure of two types of defects in FG: a F vacancy ($\text{V}_{\text{F}}$) and a double F and C vacancy ($\text{V}_{\text{CF}}$), for which different charge states were considered. 
The paper is structured as follows. 
We present the description of the theoretical method in \cref{sec:Methods}.
Our approach involves using DFT to obtain the single-particle localized states and group theory to construct the many-body configurations. 
In \cref{sec:Results} we discuss our results. 
We start with a description of pristine fluorographe and the $\text{V}_{\text{F}}$ defect in  \cref{sec:Fluorographene}. Neutral and negative $\text{V}_{\text{CF}}$ vacancies are presented  in \cref{sec:VCF,sec:VCFM}.  
We examine the transitions between ground and excited states introduced by the defects and analyze their dependence on strain. In addition, we compute the Jahn-Teller energy for the neutral defect and the zero field splitting for the negatively charged one.
Given that the symmetry of the $\text{V}_{\text{CF}}$ defect is equivalent to that of a NV center, a parallelism can be established between both systems. Based on previous NV studies, in \cref{sec:Resonators} we discuss possible applications of defective FG sheets as quantum hybrid  resonators. 
Our calculations of the formation energy of the defects are presented in \cref{sec:FormationEnergy}.
The conclusions are presented in \cref{sec:Conclusions}.   

\section{Methods \label{sec:Methods}}

The computational details of our work, based on previous studies of related 2D systems \cite{Tawfik2017,Reimers2018,Abdi2018}, are as follows. We employed the DFT code Quantum Espresso \cite{Giannozzi2009}  and used a supercell approach to study defects in FG. We used the HSE method with PBE functional \cite{PBE1996,PBE1997}, adjusting the parameter $\alpha=0.35$ to match the band gap of fluorographene obtained with the latest calculations  and experimental data \cite{Hruby2022}.  
In order to perform geometrical relaxations including HSE, we used norm conserving pseudopotentials. We used an energy cutoff of $100$ Ry and, unless otherwise stated, we used a value of $0.01$ eV/\r{A} as criterion for the convergence of the atomic forces.
We considered a $15$ \r{A} vacuum spacing between fluorographene sheets. 

For our calculations we considered $7\times7$ hexagonal supercells to avoid interaction between defects. For the calculations involving  strain in $x$ and $y$ directions we used  $7\times8$ orthogonal supercells. In both cases we considered only the $\Gamma$ point in the reciprocal space and therefore a single $q$ point in the Hartree-Fock calculation for the HSE method.  

In our study, we employed the $\Delta$SCF method \cite{Jones1989,Hellman2004} to calculate relevant transition energies, which involves computing the energy difference between the ground state and excited states with different electronic occupations. 
We determined  the vertical absorption energy (VAE) by  keeping the ground state geometry fixed  and imposing an excited electronic occupation for the calculation of the excited states. The zero-phonon line (ZPL) was  obtained after performing  a geometrical relaxation of the excited electronic configuration. 
It should be noted that the $\Delta$SCF method is applicable only to configurations corresponding to a single Slater determinant. To estimate the energy of multi-determinantal configurations, we used auxiliary single-determinant states \cite{Golami2022,MacKoit-Sinkeviciene2019}. It is worth stressing that this method provides only an estimation of the transition energies for such configurations \cite{Gali2008,Thiering2017}.

\section{Results \label{sec:Results}}
 
\subsection{Pristine fluorographene and $\text{V}_{\text{F}}$ \label{sec:Fluorographene}}
 
We obtained a lattice parameter of $2.58$ \r{A} for pristine fluorographene, in good agreement with available theoretical  \cite{Leenaerts2010,Markevich2011,Belenkov2018} and experimental \cite{Nair2010,Cheng2010} data, and a band-gap of $5.65$ eV. 

\begin{figure}[ht] 
\centering    
\subfloat{ 
  \includegraphics[height=60mm]{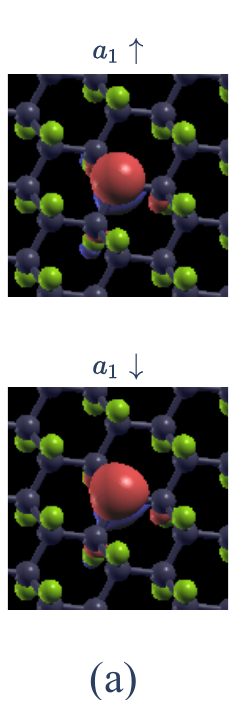} 
  \label{fig:BandStructure_VF:a} 
}
\subfloat{  
  \includegraphics[height=60mm]{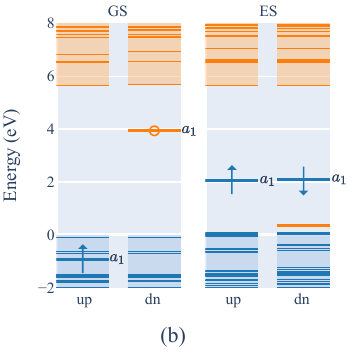}%
  \label{fig:BandStructure_VF:b} 
}   
\phantomsubfloat{\label{fig:BandStructure_VF:c}}%
\vspace{-2\baselineskip}
\caption{(a) Geometry of fluorographene with a $\text{V}_{\text{F}}$ vacancy and the localized $a_1$ orbital. 
(b) Single-particle levels for the ground state (GS) and excited state (ES) of the $\text{V}_{\text{F}}$ vacancy. The occupied (empty) single-electron states are indicated in blue (orange), and the conduction (valence) bands are shown as areas shaded with the same colors. 
The localized orbitals are labeled, and their occupancy is indicated with symbols: empty (circle) or occupied (up or down arrows). 
\label{fig:BandStructure_VF}}
\end{figure}

We start our analysis of defects with the simple fluorine vacancy, $\text{V}_{\text{F}}$,  which lowers the $C_{6v}$ symmetry of pristine fluorographene to $C_{3v}$. 
The $\text{V}_{\text{F}}$ vacancy leaves a C atom with a dangling $sp^3$ bond, which corresponds directly to a   molecular orbital (MO) with spatial symmetry $A_1$. We denoted  this single-electron orbital $a_1$. 
The geometry of the system and the $a_1$ orbital are illustrated in  \cref{fig:BandStructure_VF:a}.

According to our spin-polarized DFT calculation, the $a_1$ orbital is half occupied in the ground state (GS), resulting in a magnetic moment of the defect of $1$ and a ${}^2A_1$ many-body configuration.
The molecular orbitals of the majority (up) and minority (down) spins are well localized, with the up state located within the valence band and the down state inside the band gap (see \cref{fig:BandStructure_VF:b}). 
The first excited state (ES) can be constructed by  promoting an electron from the highest occupied valence bands, which have $E$ symmetry, to the unoccupied $a_1$ state. In this case, the well-localized $a_1$ orbital is doubly occupied, and there is a single hole in the $E$ bands. 

We calculated the VAE and ZPL  following the methodology described in  \cref{sec:Methods}, and obtained  values  of $3.44$ eV and $3.00$ eV, respectively. 
The ZPL value is consistent with absorption bands observed in less fluorinated fluorographene samples \cite{Hruby2022}. As suggested in Ref. \cite{Hruby2022}, it is likely that  the optical transitions  introduced by this midgap state were initially attributed to a much lower band-gap of fluorographene.

\subsection{$\text{V}_{\text{CF}}$  \label{sec:VCF}}

\begin{figure*}[ht] 
\centering   
\subfloat{ 
  \includegraphics[height=60mm]{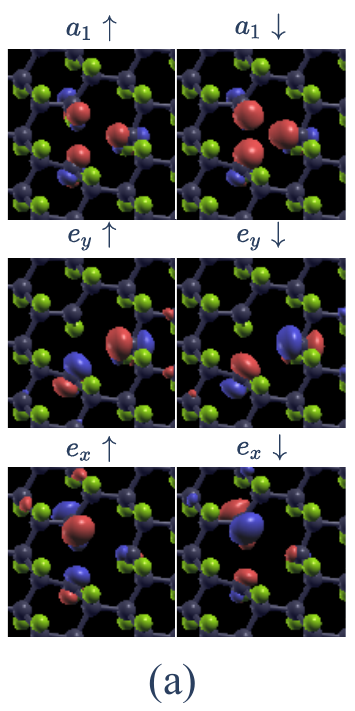} %
  \label{fig:BandStructure_VCF:a}  
}
\subfloat{  
  \includegraphics[height=60mm]{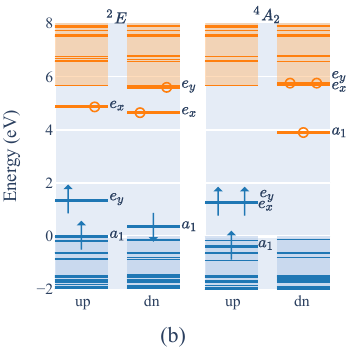} %
  \label{fig:BandStructure_VCF:b} 
}   
 \subfloat{  
  \includegraphics[height=60mm]{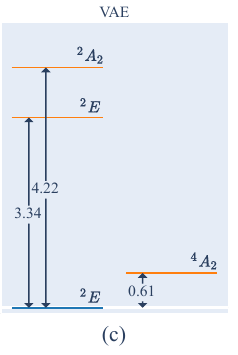} %
  \label{fig:BandStructure_VCF:c}  
}  
 \subfloat{  
  \includegraphics[height=60mm]{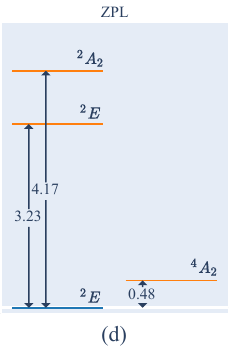} %
  \label{fig:BandStructure_VCF:d}  
} 
\caption{ (a) Geometry of fluorographene with a $\text{V}_{\text{CF}}$ defect and the resulting localized orbitals. (b) Single-particle levels for the ground state (${}^{2}E$) and first excited state (${}^{4}A_2$) for the $\text{V}_{\text{CF}}$ defect. 
The occupied and empty single-electron states are indicated with blue and orange, respectively, and the shaded areas represent the conduction and valence bands. The localized orbitals are labeled and their occupancy is indicated with symbols: empty (circle) or occupied (up or down arrows).
(c)-(d) Vertical absorption energy (VAE) and zero-phonon line (ZPL) transitions of the many-body states  referred to the ground state, in eV. The ${}^4A_2$ state is computed using the $\Delta$SCF method, while the remaining states are computed from auxiliary configurations and should be taken as estimations (see text).  Optical transitions are possible among the states in the left side of each plot, while only non-radiative transitions involve the ${}^4A_2$ state, located in the right side of the plot.  
\label{fig:BandStructure_VCF}}    
\end{figure*}

A $\text{V}_{\text{CF}}$ defect in fluorographene also lowers the symmetry of the system to $C_{3v}$. In this case, there are three $sp^3$ dangling bonds of the  C atoms around the defect, and an in-depth group-theory analysis becomes relevant.
Using the projection operator method  \cite{Tinkham2003} we determined that the three localized orbitals that can be formed have symmetries  $A_1$ and $E$. The single-particle orbitals $a_1$, $e_x$ and $e_y$ are given by 
\begin{align}
a_1&=\frac{1}{\sqrt{3}}(\sigma_1+\sigma_2+\sigma_3) \\
e_x&=\frac{1}{\sqrt{6}}(2\sigma_1-\sigma_2-\sigma_3)  \\
e_y&=\frac{1}{\sqrt2}(\sigma_2-\sigma_3) ,
\end{align}
where $\sigma_i$ corresponds to the dangling orbital of each $C$ atom. The geometry of the system and the orbitals is presented in \cref{fig:BandStructure_VCF:a}.

The most symmetric $a_1$ orbital lies lowest in energy. 
There are three electrons to fill the orbitals, so that in the ground state two electrons are located in the $a_1$ orbital, and one in an $e$ orbital. 
The configuration is then $a_1^2e^1$, and the spatial symmetry of the many-body wave function in the $C_{3v}$ symmetry induced by the defect is $A_1 \otimes A_1 \otimes E = E$. The $S=1/2$ spin of the ground state configuration gives a spin doublet, so that the total state corresponds to ${}^2E$. As discussed below, this situation is analogous to the configuration of a  neutral  $\text{NV}^0$ center \cite{Felton2008,Manson2013}. %

A neutral $\text{NV}^0$ center has four molecular orbitals formed from the corresponding dangling bonds, two with $a_1$ symmetry and a double degenerated $e$ orbital \cite{Gali2009a}. However, one $a_1$ orbital is located well below the valence band, and is not relevant for the transitions of interest. The remaining three orbitals are located within the band gap and accommodate three electrons, which is precisely the same configuration as the $\text{V}_{\text{CF}}$  vacancy in fluorographene. 
Then, the conclusions derived from group theory for  NV centers apply also to $\text{V}_{\text{CF}}$. Note that they include the resulting many-body configurations but not necessary their energy order, which is beyond a group theoretical analysis.
The similarity motivates also the study of the negatively charged $\text{V}_{\text{CF}}^-$ defect, which is analyzed in \cref{sec:VCFM}.

The many-body configurations corresponding to the ground and first excited states of the $\text{V}_{\text{CF}}$ defect are presented in \cref{tab:ECVF}. The first excited states are obtained by promoting an electron to the $E$ orbitals, that is, a $a_1^1 e^2$ configuration. The spatial symmetry of the resulting many-body states is given by $A_1 \otimes E \otimes E=A_1 \oplus A_2 \oplus E$. We constructed the electronic configurations given by the single-particle orbitals using the projection operator method. 
Note that we obtained three doublets with different symmetry and in particular a ${}^2A_2$ doublet which, as pointed out in Ref. \cite{Manson2013}, has been misidentified in some works as ${}^2A_1$ for $\text{NV}^0$.
These states can become mixed by different interactions such as spin-orbit, spin-spin, electric and magnetic fields, and strain, as analyzed for $\text{NV}^0$ in several works  \cite{Barson2019,Gali2009a,Manson2013}. 

\begin{table}[h] 
\begin{center}
\begin{tabular}{ccc}
\toprule 
${}^{2S+1}\Gamma_o$ & \text{Electronic configuration} &  \text{Label}\\
\midrule
$ ^{2}E $ & $ \ket{a_{1}\overline{a_{1}}e_{x}} , \ket{a_{1}\overline{a_{1}}e_{y}}$  &  $\mathcal{E}_{{+}1/2}^{0(x,y)}$\\ 
\midrule  
\multirow{2}{*}{$ ^{4}A_{2} $} & $ \ket{a_{1}e_{x}e_{y}}$ & $\mathcal{A}_{{+}3/2}^{1} $ \\
                               & $ \frac{1}{\sqrt{3}}\left(\ket{\overline{a_{1}}e_{x}e_{y}}+\ket{a_{1}\overline{e_{x}}e_{y}}+\ket{a_{1}e_{x}\overline{e_{y}}}\right)$ & $\mathcal{A}_{{+}1/2}^{1} $\\
\multirow{2}{*}{$ ^{2}E '$ } &  $ \frac{1}{\sqrt{2}}\left(\ket{a_{1}e_{x}\overline{e_{x}}}-\ket{a_{1}e_{y}\overline{e_{y}}}\right)  $ &  \multirow{2}{*}{$\mathcal{E}_{{+}1/2}^{2(x,y)} $ }\\
                           & $ \frac{1}{\sqrt{2}}\left(\ket{a_{1}e_{x }\overline{e_{y}}}-\ket{a_{1}\overline{e_{x}}e_{y}}\right) $ & \\
$ ^{2}A_{2} $ & $ \frac{1}{\sqrt{6}}\left(2\ket{\overline{a_{1}}e_{x}e_{y}}-\ket{a_{1}\overline{e_{x}}e_{y}}-\ket{a_{1}e_{x}\overline{e_{y}}}\right)$ & $\mathcal{A}_{{+}1/2}^{3} $ \\
$ ^{2}A_{1} $ & $ \frac{1}{\sqrt{2}}\left(\ket{a_{1}e_{x}\overline{e_{x}}}+\ket{a_{1}e_{y}\overline{e_{y}}}\right)$ & $\tilde{\mathcal{A}}_{{+}1/2}^{4} $ \\
\bottomrule
\end{tabular}  
\end{center}
\caption{Electronic configurations of the $\text{V}_{\text{CF}}$ defect. The notation ${}^2 E'$ is used to differentiate this state from the ground state with same symmetry. We label with $\tilde{\mathcal{A}}$  the state with $A_1$ symmetry, to distinguish it from the states with $A_2$ symmetry, labeled with $\mathcal{A}$. We show only the configurations with non-negative spin projections, the ones with negative spin projection can be constructed straightforwardly. \label{tab:ECVF}}
\end{table}

Given that the many-body ground state presents  spatial degeneracy, the system is Jahn-Teller unstable, giving rise to an adiabatic potential energy surface (APES) with the typical ``Mexican hat" shape \cite{Bersuker2009}.  
Therefore, the geometrical configuration of the ground state will have a symmetry lower than $C_{3v}$, namely $C_{1h}$. 
For simplicity, we keep the labels of the $C_{3v}$ symmetry for the configurations in our notation.
In our analysis, we first relaxed the system while enforcing  $C_{3v}$ symmetry to obtain the high symmetry (HS) structure. We then lifted the symmetry restriction and obtained the $C_{1h}$ lower symmetry  structure with the lowest energy (LE), a method similar to the one presented in Ref.  \cite{Zhang2018} for the study of a neutral $\text{NV}^0$ center, analogous to our system. For these calcualtions we used a stricter force convergence criterion of $1$ meV/\r{A}.

We found that the Jahn-Teller stabilization energy, which is the energy difference between the HS and LE structures, was $E_{JT}=30$ meV. This value is about one third of the value found for a neutral  $\text{NV}^0$ center  \cite{Gali2009a} and close to the one found for a negatively charged $\text{NV}^-$ center \cite{Gali2019}. 
There are three equivalent LE points separated by warping barriers, with saddle points  with an energy $\delta$ above the minimum \cite{Bersuker2009}. 
By computing the direct path between two equivalent minimum energy configurations located at different LS points, we obtained $\delta=20$ meV. 

In \cref{fig:BandStructure_VCF:b} we present the single-particle levels for both the ground state ${}^2 E$ and the first excited state  ${}^4 A_2$.
The levels $\mathcal{E}^0_{\pm1/2}$ and  $\mathcal{A}^1_{\pm3/2}$ of each manifold can be described by using a single Slater determinant (see \cref{tab:ECVF}). Therefore,  the transition energies can be obtained straightforwardly using the $\Delta$SCF method, and are given by the difference between the energies of each configuration. 
We estimated the transition energies for the remaining excited states using single  Slater determinant configurations \cite{Golami2022,MacKoit-Sinkeviciene2019,Gali2008} (see \cref{ap:multiconfigurational}). 
While this method has been successfully used to compute transition energies between multi-determinantal configurations,  it only provides a  rough estimation of the energies \cite{Gali2008}. 
For example, the method does not account accurately for the geometrical relaxation energy (Stokes shift), given that the geometry of the actual configuration cannot be computed.
In our calculations of the ZPL for the higher excited states we considered the same geometry as the one obtained for the first excited state, given that all these excited states have the same $a_1^1 e^2$ electronic occupation \cite{MacKoit-Sinkeviciene2019}. 
Note that the excited state ${}^2E'$ will also present Jahn-Teller distortion, however the accuracy of our method is not enough to estimate its $E_{JT}$. %

In \cref{fig:BandStructure_VCF:c,fig:BandStructure_VCF:d} we present the many-body states and their corresponding VAE and ZPL transition energies for the $\text{V}_{\text{CF}}$ defect. The ${}^2A_1$ state lies at $7.8$ eV and is omitted.
The values of the optical transitions from the ground state to the excited states ${}^2E'$ and ${}^2A_2$, although approximated, are consistent with available experimental data that shows absorption features at around $2.9$ eV and $4.8$ eV in less fluorinated fluorographene, attributed to single $\text{V}_{\text{F}}$ vacancies \cite{Hruby2022}. 
Only non-radiative transitions are allowed between these states and the ${}^4A_2$ state. 
The latter state is split via spin-spin interaction into two double-degenerated states, with $M_S=\pm1/2$  and  $M_S=\pm3/2$ \cite{Gali2009a}. Since the $M_S=\pm3/2$ states only couple via very weak non-axial spin-orbit interaction with the ground state, they are long-lived and have been proposed as qubit candidates for $\text{NV}^0$ centers \cite{Gali2009a}.

\begin{figure*}[ht] 
\centering   
\subfloat{ 
  \includegraphics[height=60mm]{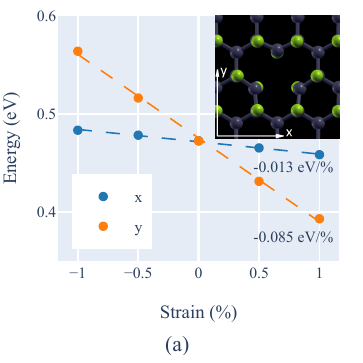} %
  \label{fig:Strain_VCF:a}  
} 
\subfloat{  
  \includegraphics[height=60mm]{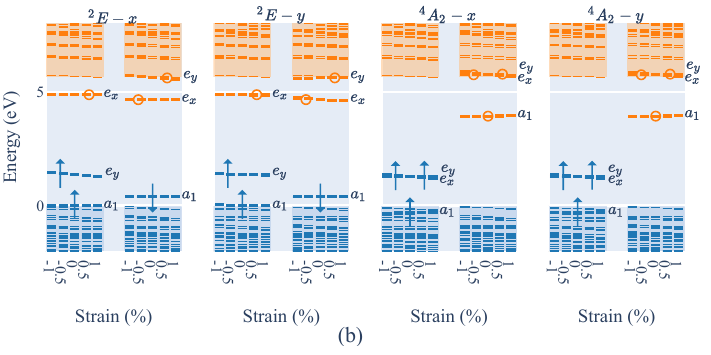} %
  \label{fig:Strain_VCF:b}   
}    
 \phantomsubfloat{\label{fig:Strain_VCF:c}}%
 \phantomsubfloat{\label{fig:Strain_VCF:d}}%
 \phantomsubfloat{\label{fig:Strain_VCF:e}}%
\vspace{-2\baselineskip}
\caption{ (a) Strain dependence of the transition energy between the ground state (${}^2E$) and the first excited state (${}^4A_2$) for a $\text{V}_{\text{CF}}$ defect, for strain applied in the $x$ and $y$ directions (shown in the inset). A linear fit is used to obtain the dependence for each direction. (b) Single-particle levels dependence with strain for each case.
\label{fig:Strain_VCF}}
\end{figure*}

In \cref{fig:Strain_VCF} we present the dependence of the ZPL transition energy between the ground state and the first excited state on strain. 
Strain is defined as the ratio of the lattice deformation ($\Delta l_{i}$) to its initial dimension ($l_{i}$), that is, $\epsilon=\Delta l_{i}/l_{i}$ with $i=x,y$.
When strain is applied in the $y$ direction, we obtain  a  variation of ${-}8.5$ eV/strain for the transition energy, whereas  we obtain  a lower value of ${-}1.3$ eV/strain when  strain is applied in the $x$ direction. 
The value in the $y$ direction is not far from the large $12$ eV/strain shift obtained for $\text{V}_{\text{N}}\text{N}_{\text{B}}$ defects in h-BN sheets \cite{Abdi2019}. 
As shown in \cref{fig:Strain_VCF:b}, strain in the $x$ direction affects mainly the $e_y$ single-particle orbital of the ground state ${}^2E$, which is occupied by an electron in our DFT calculation. 
This dependence is consistent with the geometry of the $e_y$ orbital (see \cref{fig:BandStructure_VCF:a}). 
On the other hand,  when strain is applied in the $y$ direction in the ground state, 
the occupied $e_y$ orbital remains almost constant in energy. 
Finally, in the excited state  ${}^4A_2$ both $e_x$ and $e_y$ orbitals are occupied, and the energy change when strain is applied in either direction is similar. 
As a result, when computing the energy difference between the ground and first excited states, there is a larger variation in energy when strain is applied in the $y$ direction. This is  because the energy variation of each state with strain in the $x$ direction partially compensates. 

\subsection{$\text{V}_{\text{CF}}^-$  \label{sec:VCFM}}

As discussed before, the negatively charged $\text{V}_{\text{CF}}^-$ defect possesses  the same symmetry as a $\text{NV}^-$ center. 
In \cref{tab:ECVFM} we present the many-body states corresponding to  $\text{V}_{\text{CF}}^-$,  which were obtained using   the projection technique of group theory. 
We adopt  the hole picture for the description of this defect, which is more convenient given that the electronic occupation is larger than half-filled.
The interactions arising between states  have been studied in previous works \cite{Maze2011,Doherty2011}.

\begin{table}[h] 
\begin{center}
\begin{tabular}{ccc}
\toprule 
${}^{2S+1}\Gamma_o$ & \text{Electronic configuration} & \text{Label} \\
\midrule
\multirow{2}{*}{$ ^{3}A_{2} $ } & $ \ket{e_{x}e_{y}}$ & $\mathcal{A}_{{+}1}^{0} $  \\
                                & $ \frac{1}{\sqrt{2}}\left(\ket{e_{x}\overline{e_{y}}}+\ket{\overline{e_{x}}e_{y}}\right)$ & $\mathcal{A}_{0}^{0} $\\
\midrule
$ ^{1}E $ & $ \frac{1}{\sqrt{2}}\left(\ket{e_{x}\overline{e_{x}}}-\ket{e_{y}\overline{e_{y}}}\right),\frac{1}{\sqrt{2}}\left(\ket{e_{x}\overline{e_{y}}}-\ket{e_{y}\overline{e_{x}}}\right)$ & $\mathcal{E}_{0}^{1(x,y)} $ \\
$ ^{1}A_{1} $ & $ \frac{1}{\sqrt{2}}\left(\ket{e_{x}\overline{e_{x}}}+\ket{e_{y}\overline{e_{y}}}\right)$ & $\tilde{\mathcal{A}}_{0}^{2} $ \\
\midrule
\multirow{2}{*}{$ ^{3}E $ }& $ \ket{ae_{x}},\ket{ae_{y}}$ & $\mathcal{E}_{{+}1}^{3(x,y)} $ \\
                           & $ \frac{1}{\sqrt{2}}\left(\ket{a\overline{e_{x}}}+\ket{\overline{a}e_{x}}\right),\frac{1}{\sqrt{2}}\left(\ket{a\overline{e_{y}}}+\ket{\overline{a}e_{y}}\right)$ & $\mathcal{E}_{0}^{3(x,y)} $\\
$ ^{1}E' $ & $ \frac{1}{\sqrt{2}}\left(\ket{a\overline{e_{x}}}-\ket{\overline{a}e_{x}}\right),\frac{1}{\sqrt{2}}\left(\ket{a\overline{e_{y}}}-\ket{\overline{a}e_{y}}\right)$ & $\mathcal{E}_{0}^{4(x,y)} $ \\
\bottomrule
\end{tabular}
\end{center}
\caption{Electronic configurations of the $\text{V}_{\text{CF}}^-$ defect, in the hole picture. The notation ${}^1 E'$ is used to differentiate this state from the ${}^1 E$ state with same symmetry. We label with $\tilde{\mathcal{A}}$  the state with $A_1$ symmetry, to distinguish it from the ground state with $A_2$ symmetry, labeled with $\mathcal{A}$. We show only the configurations with non-negative spin projections, the ones with negative spin projection can be constructed straightforwardly \label{tab:ECVFM}.
}
\end{table}

Only the states $\mathcal{A}^0_{\pm1}$ and  $\mathcal{E}^3_{\pm1}$ correspond to single-determinant configurations and can be calculated with the $\Delta$SCF method. 
However, the convergence of the $\mathcal{E}^3_{\pm1}$ state could not be achieved with the HSE method used. 
Note that the difficulty in convergence is expected for this case where a hole occupies a degenerated $E$ orbital ($a^1 e_x^1 e_y^2$ electron occupation) \cite{Jin2021,Gavnholt2008}. 
Then, we used the $a^1 e_x^{1.5} e_y^{1.5}$ configuration for the calculation of this state. 
By comparing the results using the $a^1 e_x^{1.5} e_y^{1.5}$ configuration with preliminary calculations using $a^1 e_x^1 e_y^2$ and a larger convergence threshold, we estimate a  difference in the energy of ${\approx}0.05$ eV, in agreement with previous reports \cite{Jin2021}.

As in \cref{sec:VCF}, we estimated the transition energies of the remaining states by using auxiliary states (see \cref{ap:multiconfigurational}). 
The excited singlets  $\mathcal{E}^1_{0}$  and $\tilde{\mathcal{A}}^{2}_{0}$ have two holes in the  orbitals with $E$ symmetry, which results in the  same electronic occupation as  the ground state.
Then, we considered the ground state geometry in the estimation of the ZPL for these excited states, assuming their ZPL energies equal to their vertical excitation energies. 
According to Hund's rules, the remaining singlet $\mathcal{E}^4_{0}$ lies higher in energy than  the excited triplet $\mathcal{E}^3_{+1}$, so that we omitted it.

\begin{figure}[ht]
\centering
\subfloat{ 
  \includegraphics[height=55mm]{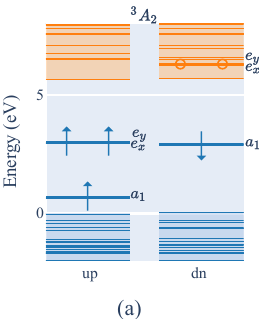} 
  \label{fig:BandStructure_VCFM:a} 
} 
 \subfloat{  
  \includegraphics[height=55mm]{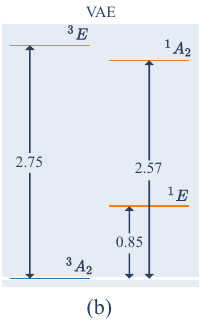} 
  \label{fig:TransitionEnergies_VCFM}  
}  \\  
\subfloat{ 
  \includegraphics[height=45mm]{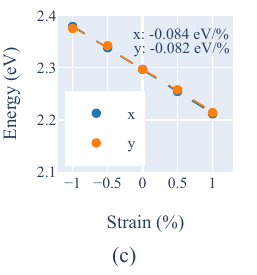} 
  \label{fig:BandStructure_VCFM:c}  
} 
\subfloat{ 
  \includegraphics[height=45mm]{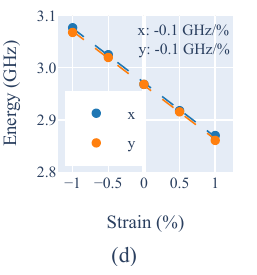} 
  \label{fig:BandStructure_VCFM:d} 
} 
\caption{(a) Single-particle levels for the ground state (${}^{3}A_2$) and the excited state ${}^{3}E$ for the $\text{V}_{\text{CF}}^-$ defect. (b) Vertical absorption energy (VAE) transitions of the many-body states  referred to the ground state, in eV. The ${}^{3}E$ state is computed using the $\Delta$SCF method, while the remaining states are computed from auxiliary configurations.  
Optical transitions are possible among states in the left or right side of the plot, while only non-radiative transitions are possible among different sides.
The estimated ZPL of ${}^{1}E$ and  ${}^{1}A_2$ are equal to the VAE. The ZPL of ${}^{3}E$ is $2.3$ eV. 
(c) Strain dependence of the transition energy between ${}^{3}A_2$ and ${}^{3}E$, for strain applied in the $x$ and $y$ directions. A linear fit is used to obtain the dependence for each direction. 
(d) Strain dependence of the ZFS of the ${}^{3}A_2$ triplet. 
\label{fig:BandStructure_VCFM}}  
\end{figure}

In \cref{fig:BandStructure_VCFM:a} we present the single-particle levels for the ground state, which can be described with a single determinant. 
The empty $e$ orbitals of the ground state ${}^{3}A_2$ are pushed up in energy into the conduction band  when compared to the same levels of the ${}^{4}A_2$ state of the neutral $\text{V}_{\text{CF}}$ defect (\cref{fig:BandStructure_VCF:b}). 
However, our DFT calculations show that these states remain well localized, and the molecular orbitals are similar to those shown in \cref{fig:BandStructure_VCF:a}. 

In \cref{fig:TransitionEnergies_VCFM} we show the VAE transitions for the excited states. As discussed before, the VAE provides an estimation of the ZPL for the singlet states. 
For  the excited triplet ${}^{3}E$ we obtained a ZPL energy of $2.3$ eV.
Note that this value of ZPL for ${}^{3}E$ is lower than the  absorption features experimentally observed \cite{Mazanek2015,Jeon2011,Hruby2022}. 
This indicates that the presence of the negatively charged defects is not energetically favored, which is consistent with the formation energy analysis presented in \cref{sec:FormationEnergy}.
Consequently, the negatively charged state should be stabilized by applying a gate voltage.

A distinguishing feature of the NV${}^-$ center defect is that it allows for high fidelity preparation of the $m=0$ sublevel of its ground state, labeled $\mathcal{A}^{0}_{0}$ in our system,  due to a convenient intersystem crossing (ISC) between triplet and singlet states \cite{Doherty2013}. 
Taking as reference the VAE of the many-body states of $\text{V}_{\text{CF}}^-$ (\cref{fig:TransitionEnergies_VCFM}), the ordering of the levels for our system would be the same as that of the NV center.
If that was indeed the case, symmetry considerations allow in principle the existence of a similar ISC, which could then be tested using available models \cite{Goldman2015}.  
However,  our rough estimations for the ZPL values suggest that the ${}^{1}A_{2}$ singlet remains above the ${}^{3}E$ triplet. 
In order to decide this question conclusively it is of considerable interest to extend this study using alternative \emph{ab-initio} methods better suited for the  calculation of multireference states \cite{Reimers2018}, since an accurate description of the states ordering is a first step to determine if an ISC similar to the one in NV${}^-$ centers is also present in $\text{V}_{\text{CF}}^-$ defects in FG.

Spin-orbit and spin-spin interactions split the excited states ${}^{3}E$ into four sublevels, and the fine structure is further split into two branches ($E_x$, $E_y$) under the application of non-axial strain \cite{Doherty2011}.
\Cref{fig:BandStructure_VCFM:c} shows the   dependence of the ZPL of the ${}^{3}E$ state with application of strain in $x$ and $y$ directions. 
We obtained a value of approximately ${-}8$ eV/strain for both directions which, as in the case of the neutral defect, is comparable to the strain shift obtained for defects in h-BN sheets \cite{Abdi2019}.

Another parameter of interest in the description of the defect is the zero-field splitting (ZFS) tensor. 
The ZFS is determined to first order by dipolar spin-spin interactions, and we calculated its value for the ground state from our DFT results \cite{Ma2020}. 
For the $C_{3v}$ symmetry of the defect, only the axial ZFS parameter $D$ is different from zero. We obtained $D=2.97$ GHz, which is close to the value for NV centers ($D=2.88$  GHz \cite{Doherty2013}).
In addition, we calculated the dependence of the ZFS for the ${}^{3}A_2$ ground state on strain (\cref{fig:BandStructure_VCFM:d}). We obtained a shift of ${-}10$ GHz/strain for both directions. 
The symmetry breaking induced by strain allows a non-zero value of the transversal component of the ZFS ($E$ parameter). 
Our calculations yield a value of $E\approx-20$ MHz for $\pm1\%$ strain, which is close to the numerical accuracy of the method used.

\subsection{Applications to hybrid  resonators  \label{sec:Resonators}}

Strain induced by the mechanical motion of  the material, for example, through the drum oscillatory modes of a FG membrane suspended from its edges,  provides an intrinsic mean of coupling phonons with electronic degrees of freedom. 
This method does not require the use of external components, resulting in a device that is less prone to noise and decoherence, and has lower complexity in its scalability than devices relying on auxiliary components to provide the coupling \cite{Lee2017}.    
However, intrinsic strain coupling is typically relatively small, which led to several proposals  aimed at increasing the interaction by using electric or magnetic fields \cite{Abdi2017}, or cavities \cite{Lee2017} coupled to the resonator.  

Depending on the system, qubits can be encoded in either the orbital or spin electronic   degrees of freedom of color centers, which makes  orbit-strain or spin-strain interactions  relevant for  phonon coupling \cite{Lee2017}. 
Typically, the spin-strain coupling strength is rather small, with values in the order of $10$ GHz/strain for  devices with implanted NV centers \cite{Ovartchaiyapong2014,Teissier2014}.
On the other hand, orbit-strain coupling is much stronger, approximately  $10^8$ times larger than spin-strain coupling, given that the molecular orbitals are directly affected by the changes in the lattice induced by mechanical motion \cite{Lee2016}. 
Values for orbit-strain coupling are typically in the range of  PHz/strain for different quantum hybrid devices using  NV centers \cite{Lee2016} and h-BN sheets with defects  \cite{Tabesh2022,Grosso2017,Abdi2019}.

The dynamics of a freestanding 2D material sheet  can be described through the elasticity theory of membranes.
In the membrane limit in which the material has vanishing thickness, which is fulfilled by single or few layers sheets, the frequency of the fundamental mechanical mode $\omega_m^0$ for membranes with simple geometries is aproximated in terms of the pretension $T$, the surface mass density $\rho_s$, a geometrical form factor $\alpha$ given by the non-trivial zero of the mode profile and a characteristic dimension of the system $d$ \cite{Wah1962,VanDerZande2010,Barton2011},
\begin{equation}
 \omega_m^0 = \frac{\alpha}{d} \sqrt{\frac{T}{\rho_s}} .
\end{equation}
For a circular membrane, $\alpha=2.4$ and $d$ is equal to the radius $R$ \cite{Wah1962,DiGiorgio2022a}, while for a ribbon of lengh $L$ clamped in the extremes, $\alpha=\pi$ and $d=L$ \cite{Tabesh2022}.
The pretension value depends on the fabrication of the membrane \cite{Belenkov2018}, 
and is related to the strain $\epsilon$ and in-plane Youngs module of the material $E_s$ by $T=E_s \epsilon$. For graphene membranes of a few $\micro$m of radius, $T$ was estimated to be ${\approx}4\times10^{-2}$ N/m \cite{Zhang2020}.
For fluorographene, $\rho_s=1.706 \text{ mg/m}^{\text{2}}$ \cite{Belenkov2018} and $E_s=100$ N/m \cite{Nair2010}.
Taking $T \approx 4\times10^{-2}$ N/m as reference, we obtain $\omega_m^0 \approx 10$  MHz for fluorographene membranes of $d\approx 1$ $\micro$m, a value in agreement with the ones obtained for similar devices of  h-BN \cite{Abdi2017} and graphene \cite{Verbiest2021,Barton2011,VanDerZande2010,Zhang2020}.
It is worth mentioning that driven devices can achieve frequencies of the order of GHz, as was obtained for $\text{MoS}_{\text{2}}$ piezo-resonators \cite{Jiang2020}.

The membrane strain is related to its deflection, and for small deflections  it can be aproximately written in terms of the maximum vertical displacement $\xi$ \cite{Zhang2020},
\begin{equation}
 \epsilon = \beta \left(\frac{\xi}{d}\right)^2 , 
\end{equation}
where $\beta$ is a geometrical factor, which for a ribbon-shaped membrane corresponds to $8/3$ \cite{Zhang2020}.
Static deflections in membrane devices can be tuned using a voltage gate, and typical values  for membranes of a few $\micro$m of radius are in the order of  $10$ nm, that leads to static built-in strains of ${\approx}10^{-4}$  \cite{Zhang2020}. 
Strain induced dynamically through time-dependent bias can achieve the same order of magnitude \cite{Zhang2020}.
The fundamental oscillation modes of micro-scale membranes around the equilibrium point have a vertical displacement of approximately $0.1$ nm, which  corresponds to an induced strain of $\epsilon \approx 10^{-8}$. These reference values correspond to a  h-BN  ribbon  \cite{Tabesh2022}. 
The quadratic dependence of the strain with the vertical displacement, which in turn depends on the membrane geometry and material through 
$\xi=\sqrt{ \hbar/ (2M\omega_m^0)}$ ($M$ is the effective mass) \cite{Tabesh2022,Abdi2017}, leads to a spread in the reference values, ranging from  $\xi\approx10^{-2}$ nm and $\epsilon \approx10^{-10}$ for a similar h-BN device \cite{Abdi2017} to $\xi\approx10$ nm and  $\epsilon \approx10^{-4}$ for the already mentioned driven resonators \cite{Zhang2020}.
For comparison, the strain of a three-dimensional (3D) diamond micro cantilever with implanted NV centers  in the fundamental mode is  ${\approx} 10^{-12}$, and can be increased to ${\approx}10^{-6}$ through mechanical drive \cite{Lee2016}. 
A scaling-down of the latter device to the nanoscale was proposed to achieve a larger orbit-strain coupling (up to the ${\approx}10$ MHz) in the fundamental mode \cite{Lee2016}, through a larger induced strain. In this regard, 2D membranes arise as promising candidates, given their comparatively large achievable strain.  

Our \emph{ab-initio} calculations suggest a deformation potential \cite{Li2020} of $\Xi \approx 1$ PHz/strain for flourographene membranes, a value similar to the one obtained for  previously studied h-BN resonators \cite{Tabesh2022,Li2020}. 
If we consider a fluorographene membrane of $d\approx 1$ $\micro$m hosting a color center and  oscillating in the fundamental mode with a vertical displacement of $\xi\approx 0.1$ nm, we obtain an orbit-strain coupling of $g=10$ MHz.

The obtained coupling is about $10^3$ times larger than the values obtained for 3D mechanical resonators with NV${}^-$ centers \cite{Lee2017,Lee2016}. 
For the latter devices, different cooling schemes were proposed \cite{Lee2017,Wilson-Rae2004,Kepesidis2013}. 
The ``off-resonant'' scheme uses the $m_s = 0$ sublevel $\mathcal{A}_0^0$ of the  ground state and the $\mathcal{E}_0^{3y}$ level of ${}^3E$ as two-level system, and convert the strain coupling to an effective transverse interaction using a laser detuned by $\omega_m^0$ from the transition energy \cite{Lee2017}. 
The ``resonant'' scheme involves  tuning  the energy difference between   the $\mathcal{E}_0^{3x}$ and $\mathcal{E}_0^{3y}$ levels of the ${}^3E$ state to be equal to $\omega_m^0$, while driving the transition from the $\mathcal{A}_0^0$ ground state to $\mathcal{E}_0^{3y}$ with a laser. This allows for resonant excitation to the $\mathcal{E}_0^{3x}$ state by removing a phonon from the mechanical mode \cite{Lee2017}. 
However, scaling down these devices from the microscale to the nanoscale is necessary to achieve ground-state cooling using these methods \cite{Lee2017,Lee2016}.
The inherent larger coupling in our system would enable the implementation of these methods in a flourographene membrane device,  thereby extending   the proposal for  the NV${}^-$ center to the $\text{V}_{\text{CF}}^-$ defect. 
Another possible protocol uses the $\mathcal{A}_{\pm1}^{0}$ levels in a $\Lambda$ configuration with an excited state formed with the $\mathcal{E}_{\pm1}^{3(x,y)}$ levels, which are mixed through spin-orbit interaction \cite{Doherty2011}. This scheme relies on stimulated Raman transitions to remove phonons from the resonator, and has the advantage of combining the stronger orbit-strain coupling with the  larger coherence of spin states \cite{Lee2017}.

\subsection{Formation energy and stability \label{sec:FormationEnergy}}

The formation energy for a defect with charge $q$ is obtained from \cite{Freysoldt2014a} %
\begin{equation}\label{eq:FormationEnergy}
 E_f^q(\epsilon_F)=E_{d}^q-E_{bulk}-\sum_i n_i \mu_i +q \epsilon_F + E_{corr}^q ,
\end{equation}
where $E_{d}^q$ is the total energy of the supercell with the defect, $E_{bulk}$ is the energy of the pristine supercell, $n_i$ are the number of atoms that have been added ($n_i>0$) or removed ($n_i<0$) to form the defect, with $\mu_i$ the corresponding chemical potentials. 
The energy depends on the total charge with the Fermi energy $\epsilon_F$, measured from the top of the valence band. 
The final term $E_{corr}^q$ accounts for  corrections such as finite $\mathbf{k}$-point sampling and electrostatic interactions \cite{Freysoldt2014a,Gali2023}. Here, we apply the Freysoldt–Neugebauer–Van de Walle (FNV) correction scheme \cite{Freysoldt2009,Naik2018}.

\begin{figure}[ht]
\centering
\includegraphics{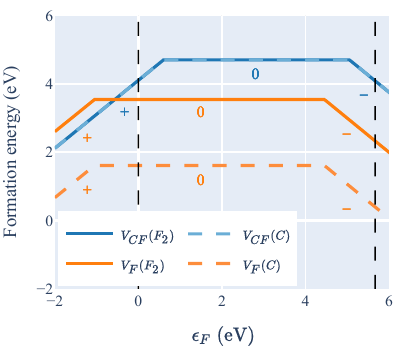} %
\caption{Formation energy as a function of the Fermi energy for $\text{V}_{\text{F}}$ and $\text{V}_{\text{CF}}$ defects, for F${}_2$-rich and C-rich environments. The black dashed lines indicate the position of the top of the valence and the bottom of the conduction bands. The favorable charge states ($+$, $0$ or $-$) of the defects are also indicated. 
\label{fig:FormationEnergies}} 
\end{figure}

In \cref{fig:FormationEnergies} we present the formation energy for the $\text{V}_{\text{F}}$ and $\text{V}_{\text{CF}}$ defects as a function of the Fermi energy, which can be varied by applying a gate voltage. 
We considered two different scenarios for the chemical potentials. 
In the first scenario, the defective membrane is in equilibrium with F${}_2$, which results in a fluorine-rich environment. For this case, we obtain $\mu_{\text{F}}=\mu_{\text{F}_2}/2$ from the energy of a F${}_2$ molecule and $\mu_{\text{C}}=\mu_{\text{FG}}-\mu_{\text{F}}$ from the difference between  the energy of the pristine fluorographene primitive cell ($\mu_{\text{FG}}$) and the fluorine chemical potential. 
In the second scenario, we  considered a carbon-rich environment and calculated $\mu_{\text{C}}$ from a graphene primitive cell. We obtained  the fluorine chemical potential from the difference with $\mu_{\text{FG}}$, which gives $\mu_{\text{F}}=\mu_{\text{FG}}-\mu_{\text{C}}$.

The formation energy of $\text{V}_{\text{CF}}$ is independent from the environment, since the third term in \cref{eq:FormationEnergy}, $\sum_i n_i \mu_i=n_{\text{C}} \mu_{\text{C}} +n_{\text{F}} \mu_{\text{F}}$,  equals $\mu_{\text{FG}} $ by definition for both environments. On the other hand, the formation energy for the $\text{V}_{\text{F}}$ defect is higher in the F${}_2$-rich environment, as expected. 
The formation energy of $\text{V}_{\text{F}}$ is higher than that of $\text{V}_{\text{CF}}$ only in the special condition of F${}_2$-rich environment and $\epsilon_F \lesssim -0.5$ eV. For the remaining conditions, the $\text{V}_{\text{F}}$ defect is more stable than $\text{V}_{\text{CF}}$. However, molecular dynamics calculations suggest that the latter defect is also thermodynamically stable  \cite{Li2021}.

\section{Conclusions \label{sec:Conclusions}}

In this study, we investigated the electronic properties of $\text{V}_{\text{F}}$,  $\text{V}_{\text{CF}}$  and $\text{V}_{\text{CF}}^-$ defects in FG membranes.
We computed the many-body states from single-particle DFT results making use of group-theoretical considerations, obtained the transition energies between the states and analyzed their dependence with non-axial strain. 
The obtained energy shift under strain for the studied defects was in the order of $1$ PHz/strain, which is comparable to the one found for defects in h-BN sheets. 
This value leads to an orbit-strain coupling of $g\approx10$ MHz for membranes of ${\approx} 1$ $\micro$m. 
Due to the similarities of $\text{V}_{\text{CF}}$ defects in FG with NV centers on diamond, some proposals for NV centers resonators can be mapped to 2D devices based on FG with $\text{V}_{\text{CF}}^-$ defects, taking advantage of the larger strain achievable in 2D materials.
Furthermore, extending this study with alternative \emph{ab-initio} methods would be useful to determine if an ISC similar to the one present in NV centers could also be expected in this system.
Our findings suggest that the $\text{V}_{\text{CF}}$ defect in FG membranes can be a promising candidate for developing nanomechanical resonators with strong orbit-strain coupling and contribute to the understanding of defects in two-dimensional materials and their quantum applications.

\vspace{10pt}

\begin{acknowledgments}

This work was supported by the ERC Synergy grant
HyperQ (Grant No. 856432) and by the BMBF via the
project CoGeQ (grant No. 13N16101). The authors acknowledge support by the state of Baden-Württemberg through bwHPC and the German Research Foundation (DFG) through grant No. INST 40/575-1 FUGG (JUSTUS 2 cluster).
M. S. T. thanks J. S. Pedernales for helpful discussions. 

\end{acknowledgments}

\appendix

\section{Multi-configurational States \label{ap:multiconfigurational}}
       
For $\text{V}_{\text{CF}}$, the ground  and first excited states are directly described by a single determinant, while the remaining states are multi-configurational.  
In principle, the $\Delta$SCF method does not allow to compute configurations composed by several determinants. However, it is possible to obtain a rough estimation of these multi-configurational states from  single-determinant auxiliary configurations  \cite{Golami2022,MacKoit-Sinkeviciene2019}. 
To illustrate the method, consider the state $\mathcal{A}^{3}$. We note that 
\begin{equation}
 \sqrt{2} \mathcal{A}_{+1/2}^{3}  + \mathcal{A}_{+1/2}^{1} =\sqrt{3} \ket{\overline{a_{1}}e_{x}e_{y}}  .
\end{equation} 
Considering that the energy of the states is independent of the spin projection, $E( \mathcal{A}_{+1/2}^{1} )=E( \mathcal{A}_{+3/2}^{1} )=E( \mathcal{A}^{1} )$, we obtain
\begin{equation}
E( \mathcal{A}^{3} )=\frac{1}{2}(3E( \ket{\overline{a_{1}}e_{x}e_{y}}  )-E( \mathcal{A}^{1} )). 
\end{equation}  

Similarly, for the two remaining excited states of $\text{V}_{\text{CF}}$ we obtain the following expressions, 
\begin{align}
E(\mathcal{E}^{2})&=\frac{1}{3}\left(6E(\ket{a_{1}e_{x}\overline{e_{y}}})-2E( \mathcal{A}^{1} )-E(\mathcal{A}^{3})\right) \\
E(\mathcal{A}^{4} )&=2E(\ket{a_{1}e_{x}\overline{e_{x}}} )-E(\mathcal{E}^{2}) .
\end{align} 

For $\text{V}_{\text{CF}}^-$, we obtain the following expressions for the  transition energies of the multiconfigurational states 
\begin{align}
 E(\mathcal{E}^1)&=2E(\ket{e_{x}\overline{e_{y}}})  \\
 E(\mathcal{A}^{'2})&=2 E(\ket{e_{x}\overline{e_{x}}})-E(\mathcal{E}^1)   \\
 E(\mathcal{E}^4)&=2E(\ket{a\overline{e_{x}}}) .
\end{align}

\bibliography{Main}


\end{document}